# Phase transformation and water adsorption behavior of ALD deposited and annealed Ru and RuO$_2$ films.

S. S. Nalawade[1], R.S. Kim[3], J. Mahl[3], S. Cherono[2], I. Chris-Okoro[2], V. Craciun[6,7], J. Yan[4], E. Crumlin[3,5], D. Kumar[2], S. Aravamudhan[1*]

[1]Department of Nanoengineering, North Carolina A & T State University, Greensboro, North Carolina, United States

[2]Department of Mechanical Engineering, North Carolina A & T State University, Greensboro, North Carolina, United States

[3]Chemical Sciences Division, Lawrence Berkeley National Lab, Berkeley, California, United States

[4]Molecular Biophysics and Integrated Bioimaging Division, Lawrence Berkeley National Lab, Berkeley, California, United States

[5]Advanced Light Source, Lawrence Berkeley National Lab, Berkeley, California, United States

[6]National Institution for Laser, Plasma and Radiation Physics, Magurele City, Romania

[7]Extreme Light Infrastructure for Nuclear Physics, Magurele, Romania

*Email: saravamu@ncat.edu

**Abstract**

Ruthenium metal and its oxide stand out for their exceptional catalytic activity, stability in Oxygen Evolution Reactions (OER) and electrical conductivity, making them indispensable in electronics and electrocatalysis. In this study, atomic layer deposition (ALD) was used to synthesize ruthenium thin films, and the subsequent annealing of deposited ruthenium films at different elevated temperatures resulted in a progressive phase transformation from ruthenium metal to ruthenium dioxide (RuO$_2$). The films were systematically characterized using atomic force microscopy (AFM), X-ray photoelectron spectroscopy (XPS), X-ray diffraction (XRD), and Raman spectroscopy. XPS was carried out with both soft X-rays from a lab-based instrument and tender X-rays from the synchrotron. The different probing depths of the techniques revealed the gradual transformation of Ru to RuO$_2$ from the top surface as the annealing temperature was increased. The water adsorption behavior of the films was also assessed using ambient pressure XPS (APXPS) at different water vapor pressures. The influence of annealing conditions on the films' affinity for water and tendency for water dissociation was analyzed to seek an initial understanding of the surface chemistry relevant to electrochemical water splitting.

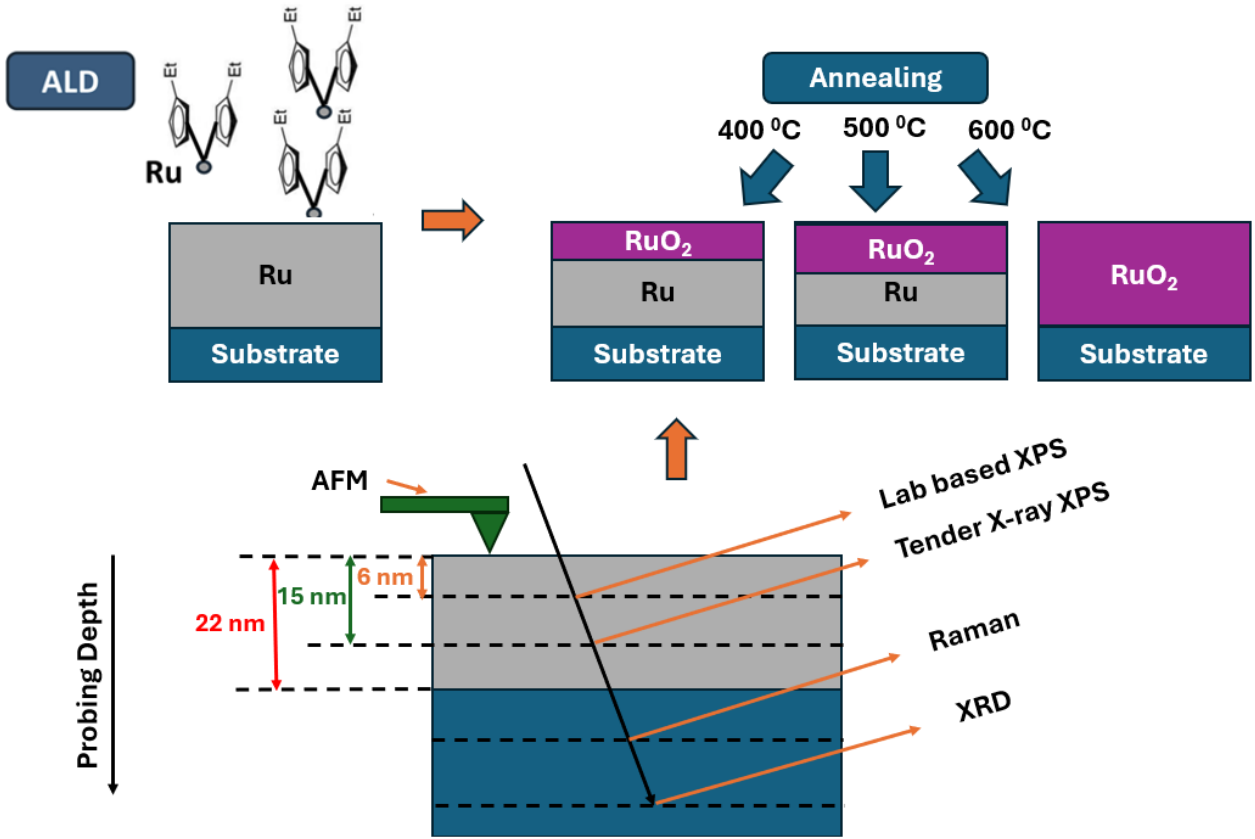

# 1. Introduction

Ruthenium (Ru$^0$) is a rare transition metal known for its hardness, high resistance to corrosion, good thermal, chemical stability, low resistivity (bulk resistivity of Ru metal ~ 7.4 µΩ.cm) and excellent catalytic capabilities. These properties make it highly valuable in various industrial applications, including electrical interconnects and electrochemical applications .[1-3] Additionally, ruthenium is employed as a catalyst in chemical reactions such as production of ammonia, CO oxidation and plays a critical role in advanced photovoltaic technologies for solar energy harvesting.[4,5] Ruthenium Dioxide (RuO$_2$) on the other hand, is a naturally highly conductive (bulk resistivity ~ 46 µΩ.cm) metallic oxide that is notably used for its outstanding catalytic activity, particularly in electrochemical applications such as supercapacitors and water splitting for hydrogen production.[6,7,8] It serves as an effective catalyst for the OER, a key process in electrochemical energy conversion and storage systems due to its exceptional stability in both acidic and basic media.[9] The combination of high conductivity and catalytic efficiency makes RuO$_2$ an exceptional material in the field of electrocatalysis and electronics.[10] For these applications, well defined smooth and pinhole-free surfaces of ruthenium and ruthenium dioxide thin films are synthesized using a variety of deposition techniques such as physical vapor deposition, chemical vapor deposition, molecular beam epitaxy, pulsed laser deposition and atomic layer deposition (ALD).[11-16] In recent years, there has been considerable focus on depositing ruthenium and ruthenium dioxide in electrochemical applications through ALD as a function of the deposition conditions such as growth temperature and oxidant exposure duration. [50,51] Due to recent development in the semiconductor processing, ALD has got more attention and International Technology

Roadmap for Semiconductors (ITRS) has included ALD for high dielectric constant gate oxides in the MOSFET structure and for copper diffusion interconnects.[17] ALD enables the precise formation of ultra-thin films, conformal passivation layers, ensuring effective protection against environmental degradation and surface contamination.[18] These layers are particularly valuable in enhancing the performance and reliability of electronic and photovoltaic devices. ALD is suitable for the deposition of high-performance transition metals and its oxides for different applications due to several distinct advantages such as atomic scale thickness control, uniformity and conformality, precise control of oxidation states and low processing temperature.[19] The phase of the ruthenium is controlled by the deposition temperature and pulsing time of the co-reactant. It had been found that higher temperature between 290 to 320 $^0$C is favorable for the ruthenium metallic deposition and lower temperature window between 220 to 240 $^0$C is best for the deposition of $RuO_2$.[20] Similarly, dosing of oxygen pulse had been studied as deciding factor of the ruthenium metallic phase. The chemical reaction for the growth of Ru metal with main by-products of water and carbon dioxide can be suggested as: [21]

$$\text{Ru (EtCp)}_{2 \text{ (gas)}} + \frac{37}{2} \text{O}_{2 \text{ (gas)}} \rightarrow \text{Ru (solid)} + 9\text{H}_2\text{O} + 14\text{ CO}_{2 \text{ (gas)}} \quad (1)$$

Lowering the deposition temperature and increasing the oxygen pulse duration had resulted in deposition of $RuO_2$ thin films. [20,22,23] Kim et al. also studied the effect of ruthenium exposure time and oxygen partial pressure in the oxygen pulse. An elevation in the $O_2$ flow rate led to increased growth rates and resistivity, attributed to the formation of $RuO_2$ films. This outcome stemmed from prolonged oxygen pulses inducing a phase transition from Ru to $RuO_2$, wherein oxygen gets adsorbed onto the surface following the combustion of ligands during the ruthenium layer formation.[24] The effect of the pulsing time ratio of oxygen to ruthenium showed that after fixing process pressure to 1 Torr, the ratio > 10 exhibited RuO$_2$ films. The growth rate of RuO$_2$ had been observed to be independent of the ruthenium pulse after self-limiting nature was achieved but showed some dependence on the oxygen dose at high doses. After the initial incubation delay, the film thickness behaves as a linear function of number of deposition cycles.[25] Delayed nucleation, characterized by extended incubation cycles, results in a rough film morphology that is suboptimal for achieving conformal deposition on structures with high aspect ratios. Recent investigations by Park et al showed that the nucleation problem of ALD ruthenium could be addressed by using several zero valent precursors compared to the precursors with higher metal valences.[26] Diluting the oxygen with Argon carrier gas throughout the ALD steps resulted in the increase in growth per cycles (GPC) of the polycrystalline RuO$_2$ compared to conventional ALD of ruthenium with excellent step coverage exceeding 80 %.[27] However, these zero valent precursors possess lower thermal stability, needs higher precursor dosing time and causes formation of less stable oxidation state of ruthenium like RuO$_3$ as compared to RuO$_2$. Although diluted oxygen improved the growth per cycle (GPC) of RuO$_2$, it directly contributed to an increase in film resistivity, which could negatively affect its performance in electrocatalytic applications. Highly stable and conductive RuO$_2$ films were achieved by optimizing the dosing time of Ru(EtCp)$_2$ in combination with pure oxygen as the reactant.

In this work, we investigated the structural, morphological, and chemical properties of metal ruthenium thin films deposited by ALD at 300°C, and examined how post-annealing influences these properties. Annealing progressively transformed the phase of the as-deposited Ru metal to highly active catalyst rutile RuO$_2$ for electrochemical applications.

The chemical properties as a result of the phase change are confirmed with lab based XPS and synchrotron tender X ray photoelectron spectroscopy. XRD showed the phase transformation from as-deposited Ru metal hexagonal closed packed to rutile $RuO_2$ at 600 $^0C$. APXPS study revealed that there is progressive change in the oxidation state with the temperature and this change in oxidation enhanced water adsorption which is crucial for electrochemical water splitting applications.

## 2. Experimental Methods

The deposition of Ru thin films on a single crystal $TiO_2(110)$ substrate was carried out using a liquid $Ru(EthylCycloPentaDienyl)_2$ ($Ru(EtCp)_2$) ruthenium precursor (Strem chemicals Inc) and oxygen as the co-reactant. The deposition process was completed in a Veeco Savannah S200 reactor with a chamber diameter of 25 cm. The chamber was heated at a temperature of 300 $^0C$. The organometallic compound $Ru(Etcp)_2$ was heated up to a temperature of 85 $^0C$ to create enough vapor pressure and introduced into a reaction chamber using ultrapure $N_2$ as the carrier gas. The flow rate for $N_2$ was 50 sccm throughout the deposition process. The ALD cycle comprised a 1-second pulse of ruthenium, followed by 10 seconds of nitrogen purging, 0.05 seconds of oxygen pulse, and 10 seconds of nitrogen purging. The substrates were cleaned using piranha solution which was a mixture of sulfuric acid and hydrogen peroxide in a ratio of 3:1, for 15 minutes and then sonicated in DI water for 5 minutes. The oxygen plasma treatment was done on cleaned substrates to remove residual carbon contamination using plasma etch system for 5 minutes with RF power of 60 W.

In this study, surface morphology was probed using an Asylum Research MFP 3D Atomic Force Microscopy (AFM) was used in AC air topography tapping mode. A pyramidal, antimony (n) doped silicon cantilever tip with backside coating of reflective aluminum and resonant frequency 320 kHz (127 μm length, 35 μm width, 42 N/m spring constant) was used to acquire the surface topography and roughness of the samples. During the analysis, images were collected over 1 μm × 1 μm area at scan rate of 0.8 Hz with a 256 × 256 pixels image resolution to ensure detailed surface feature capture. To reveal the bulk phase of the grown films we utilized X-ray diffraction (XRD), and X-ray reflectivity (XRR) measurements were carried out using a Rigaku Ultima IV diffractometer equipped with a Cu Kα radiation source (λ = 1.5418 Å) and a voltage of 40 kV. The films were scanned in a 2θ range from $20^0$ to $70^0$ at a scan speed of $4^0$ per minute. Lab based X-ray photoelectron spectroscopy (XPS) were carried out using Thermo Scientific ESCALAB$^{TM}$. A 500 μm-diameter beam with monochromatic Al Kα x-rays (1486.6 eV) is used. Tender X- ray APXPS at Beamline 9.3.1 at the Advanced Light Source (ALS), Lawrence Berkeley National Laboratory (LBNL) with photon energy of 4000 eV was utilized to investigate ~ 15 nm into the film to understand the elemental composition and surface chemistry of as-deposited Ru metal films and annealed films in vacuum as well as exposed to different water vapor conditions ranging from UHV to 17 Torr which is approaching 100% relative humidity at room temperature. The analyzer is a Scienta Omicron R4000 HiPP-2 with an energy resolution of 250 meV. Raman measurements were performed using Witec Alpha Confocal Raman spectrometer equipped with a 532 nm Nd: YAG laser. The laser power was kept below 10 mW to prevent damage to thin films. Spectra were collected in the range of 100 to 800 cm$^{-1}$ with a spectral resolution of 1 cm$^{-1}$. Accumulation and integration time were 1 second and 100 seconds respectively to improve signal noise ratio. Samples were analyzed at room temperature under ambient conditions. The post annealing of the films was performed in a Thermo Fisher tube furnace in the open air for 1.5 hours at elevated temperatures of 400 $^0$C, 500 $^0$C and 600 $^0$C. The as-deposited sample is represented as S1 asdep, sample annealed at 400 $^0$C is S2, sample annealed at 500 $^0$C is S3 and sample annealed at 600 $^0$C is S4. We will use this nomenclature of the samples throughout the discussion.

3. Results and Discussion

3.1 ALD growth of Ru metal

We first characterized the ALD growth of as-deposited Ru film with AFM in AC air topography mode to obtain high magnification images of the as-deposited Ru thin metal film. The RMS surface roughness was around 1 nm indicated smooth surface.

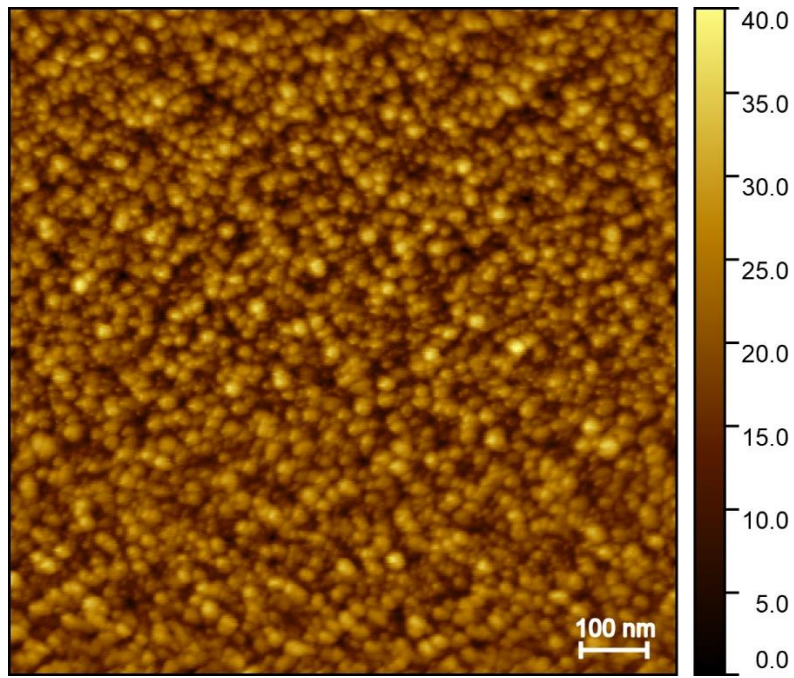

**Figure 1. Atomic force microscopy (AFM) image of as-deposited Ru metal film (Sample S1).**

**Figure 1** shows 2D imaging of as-deposited Ru (S1 asdep) which reveals distinct, well-defined grains. The scan area was 1 μm × 1 μm. The RMS surface roughness was found to be 4.8 nm which verifies the superior quality of the film deposition using ALD. [28] X-ray diffraction (XRD) was performed to determine the crystalline structure of as deposited ruthenium thin films at 300 $^0$C. The XRD pattern revealed prominent diffraction peaks (red color) at 2-theta values of 38.45$^0$, 42.79$^0$ and 44.09$^0$ as shown in **Figure 2**. These peaks were indexed to the (100), (002) and (101) planes respectively, of the hexagonal close packed (hcp) structure of ruthenium metal. [29,30] The close alignment of these observed peaks with the standard diffraction data confirmed the formation of a highly polycrystalline ruthenium metal phase in the films. There was no peak from $RuO_2$ crystal orientation even though the co-reactant was pure oxygen gas which suggests that the deposition conditions at 300 $^0$C favor the formation of ruthenium metal, where the activation barrier for oxidation of ruthenium-to-ruthenium oxide is still not satisfied that is typically seen at lower temperatures. [32] The XRR data of S1 as-dep revealed a film thickness of around 22 nm. The estimated density of as deposited sample S1 was determined to be 12.36 g/cm$^3$, which matches the known bulk density of Ru metal (theoretical density of Ru ~ 12.37 g.cm$^{-3}$). This finding showed that the as-deposited Ru metal ( sample S1) films were dense and well packed. Such a high density is indicative of minimal porosity and high atomic packing efficiency. XRR measurements have shown a highly consistent deposition process. This level of uniformity is critical for the reliable performance of the deposited Ru thin films.

XRD measurements assure the deposition of Ru metal films with sharp peaks of hexagonal crystal structure.

The XPS spectrum of the as-deposited Ru sample predominantly exhibited characteristics of Ru metal ($Ru^0$). Notably, the $Ru3d_{5/2}$ and $Ru3d_{3/2}$ (red color) core level peaks were observed at 279.9 eV and 284 eV respectively as shown in **figure 3 (a)**.[33,34] These binding energies are consistent with the presence of metal Ru in a metallic $Ru^{(0)}$ state. The spin orbit splitting between core levels is observed to be 4.1 eV which is consistent with Ru metal films.[35] There was a feature emerged at 280.6 eV, assigned to $Ru^{+4}$ which suggested the oxidation of the top film to amorphous $RuO_2$. However, the peak intensity of $Ru^{+4}$ was much less than that of peak intensity of $Ru^0$. The electronic structure of ruthenium causes a characteristic satellite peak for Ru $3d_{5/2}$ and $Ru3d_{3/2}$ orbitals corresponding to shake up phenomena of transition metals and their oxides.

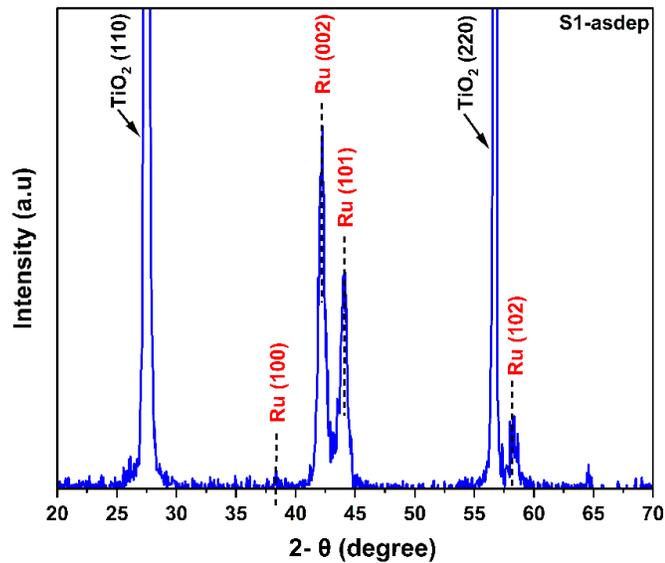

**Figure 2. XRD spectrum of as-deposited Ru metal film (sample S1-asdep)**

The Ru(3d) core level was fitted with an asymmetric line shape, a common approach for transition metals and their conductive oxides.[37] The satellite peaks (sat.1 and sat.2) (blue color) associated with Ru3d$_{5/2}$ and Ru3d$_{3/2}$ are observed at 281.8 eV and 285.9 eV respectively which are generally ~ 1.8 to 2 eV above its core level. Due to strong overlap between Ru3d$_{3/2}$ and C1s orbitals, it is difficult to quantify carbon region, however with fixed constraints on the fitting algorithm including FWHM and keeping spin splitting constant to 4.1 eV gave well fitted peaks for C1s (olive color) at 284.8 eV.[37] The metal carbonate species are assigned to the peak observed at 288 eV. The peak at 290 eV is from the organic C=O species which tells some surface carbon contamination is in the oxidized form.

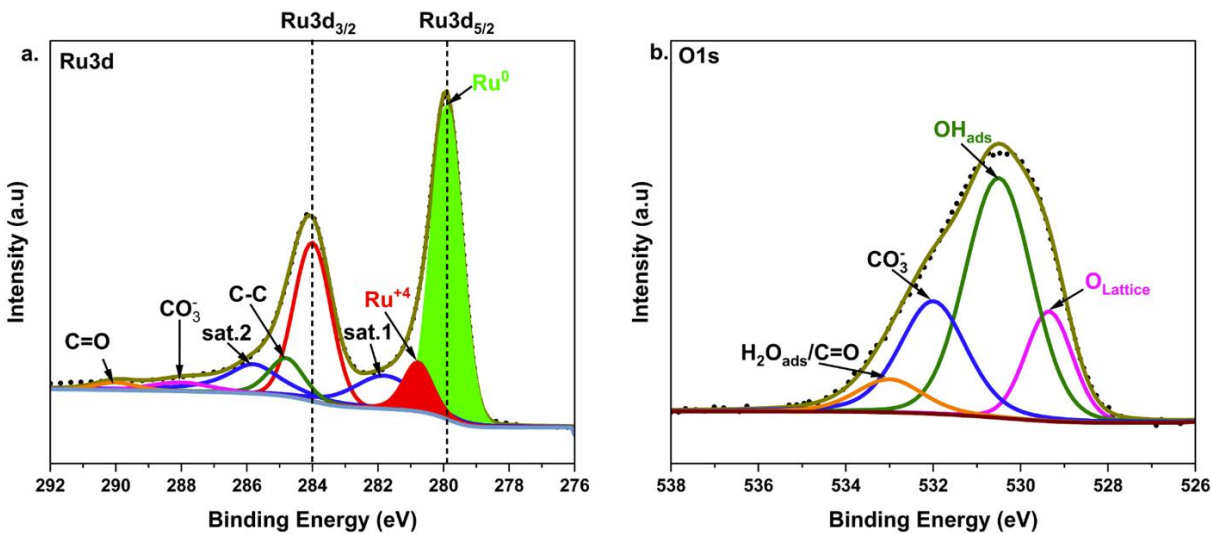

**Figure 3. Core level (a) Ru3d and (b) O1s spectra of as deposited Ru metal sample (S1 asdep) from lab based XPS**

The deconvoluted oxygen spectrum in Figure 3(b) reveals a peak at 529.3 eV, suggesting the formation of a native surface oxide on the top of the Ru metal film. The adsorbed OH species are observed at a binding energy of 530.7 eV, $CO_3^-$ species at 532 eV. The components due to water adsorption and organic carbon contamination are overlapped with each other at 533 eV, so this peak is assigned to $H_2O_{ads}$ and C=O. These peaks at 533 eV are due to exposure of the sample to moisture and carbon, respectively in the air.[20] The fitting parameters for O1s deconvolution includes the FWHM in the range of 1.5 ± 0.1 eV for -$OH_{ads}$, $CO_3^-$ and $H_2O_{ads}$/C=O and that for $O_{Lattice}$ in the range of 1 to 1.2 eV. [41] The Raman spectrum of as deposited Ru films is shown in **figure 4**. The vibrational modes for the as-deposited sample S1 were observed at 189 $cm^{-1}$ with sharp peak, other peaks at 235 $cm^{-1}$, 447 $cm^{-1}$ and 611 $cm^{-1}$ were observed and attributed to the $TiO_2$ (110) substrate. The peak at 189 $cm^{-1}$ corresponds to ruthenium-ruthenium vibrations in the metallic lattice. In some cases, peaks in this low wave number range could be associated with the phonon modes of the metal.

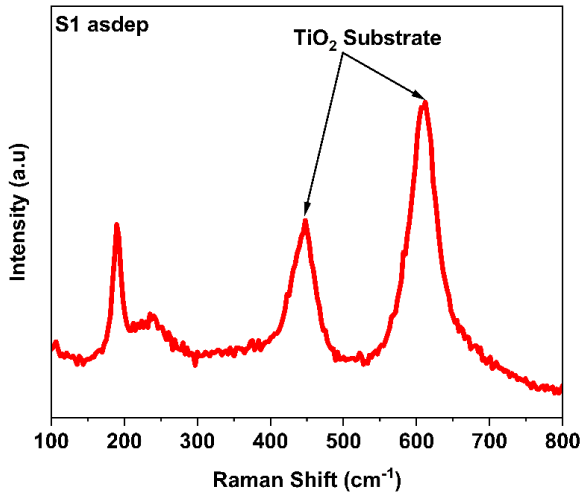

Figure 4. Raman spectra of as-deposited Ruthenium at 300 $^0$C.

**3.2 Impact of annealing on Ru metal films.**

The as-deposited sample S1 has shown the Ru metal phase with low oxygen which was seen in the XPS data and metal phase in the bulk was confirmed by hexagonal closed packed orientation in XRD. These samples were annealed at elevated temperatures of 400 $^0$C, 500 $^0$C and 600 $^0$C in open air in a tube furnace for 1.5 hours with a ramping rate of 2 $^0$C/min. **Figure 5** (a), (b) and (c) shows AFM images of the samples annealed at 400 $^0$C, 500 $^0$C and 600 $^0$C respectively. The scale of all the images were kept the same for better visual comparison and consistency in surface roughness measurements. The as deposited sample S1 ( RMS roughness ~ 1 nm), when annealed to 400 $^0$C (sample S2), showed increase in the RMS surface roughness to 2.1 nm and the enhanced grain size. This could be attributed to enhanced atomic mobility and diffusion at elevated temperatures, promoting grain growth while increasing surface unevenness due to differential growth rates as some grains may grow faster than others due to their favorable orientation or higher atomic mobility and possible contributions from oxidation.[30,34]

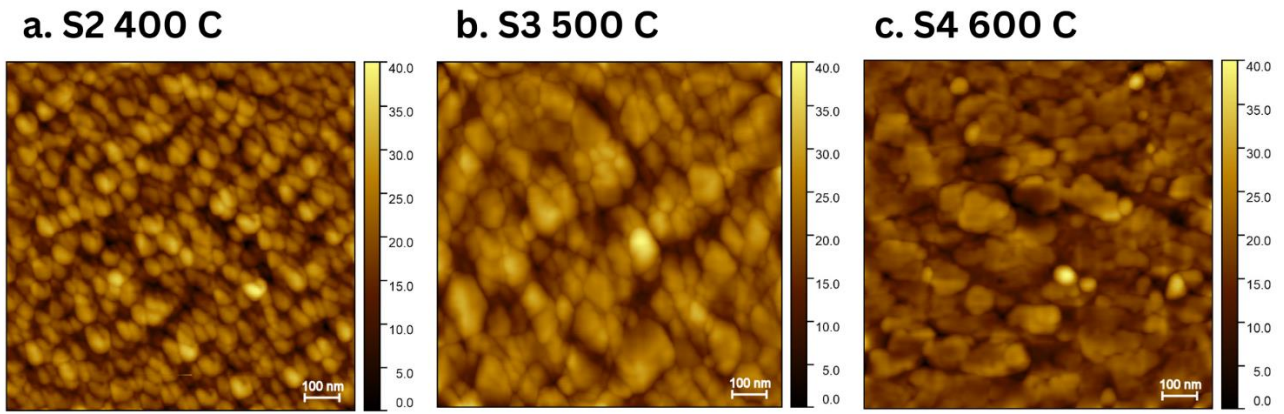

Figure 5. 2D AFM and RMS surface roughness of (a) S2, (b) S3 and (c) S4

When as-deposited sample annealed to 500 °C, its surface roughness increased, and similar to annealing temperature at 600 °C. **Table 1** shows increase in AFM statistical parameters such as RMS surface roughness.

**Table 1. AFM statistical parameters of S1 asdep, S2, S3, S4**

| Sample | RMS Roughness (nm)[a] |
|---|---|
| S1 asdep | 1 |
| S2 400 | 2.1 |
| S3 500 | 2.54 |
| S4 600 | 4.37 |

a) Nanometer

**Figure 6 (a), (b) and (c)** shows XRD spectrum of the sample S2 annealed at 400 °C, sample S2 annealed at 500 °C and sample S4 annealed at 600 °C. It is important to note that XRD is a bulk characterization technique providing crystallinity information mainly from the entire sample rather than focusing on surface specific details. The as-deposited sample S1 showed pure metal hexagonal crystal structure. This might be attributed to less oxygen incorporated into metal lattice of as-deposited S1 and it suggested that for complete phase transformation of Ru metal, there would be requirement of higher annealing temperature. This was also seen in XPS where there was still sub stoichiometric proportion of Ru, oxygen in the film and binding energy was associated with $Ru^{(0)}$ highest intensity. The annealed sample S2 at 400 °C showed metal Ru structure (red color) with preferential orientation in (002), (101) and (102) direction with 2θ at $42.2^0$, $44.04^0$ and $58.4^0$ respectively.[20,25] The sample S3 annealed at 500 °C showed up a peak related to $RuO_2$ in

(220) (magenta color) orientation seen at 57.9⁰.[24] There were still presence of peaks related to hexagonal closed packed structure of Ru metal which showed S3 had mixed phase of Ru metal as well as RuO$_2$.

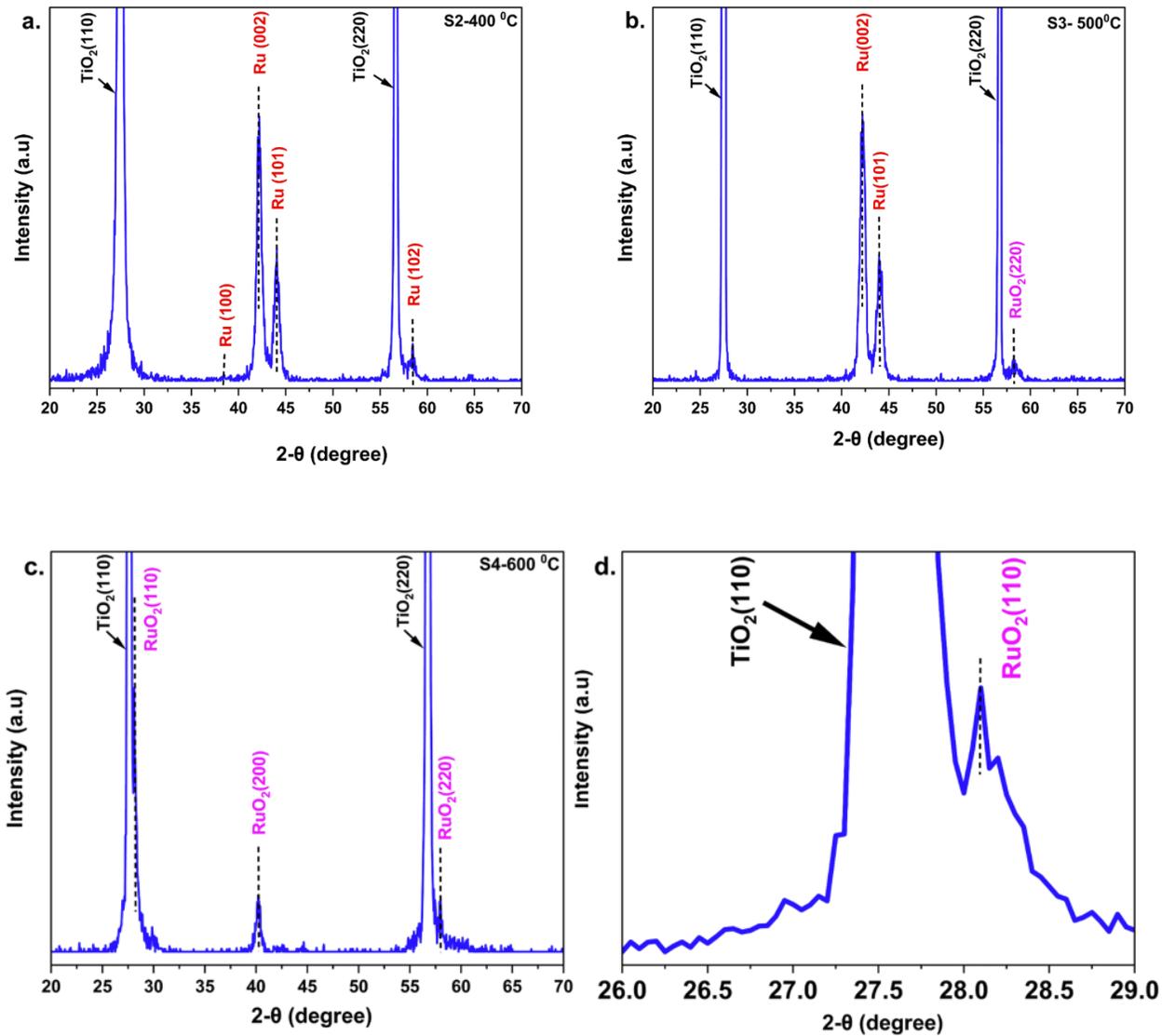

**Figure 6. XRD analysis of a)annealed sample S2, b)annealed sample S3, c) annealed sample S4 and d) zoomed image of annealed sample S4 showing rutile RuO$_2$(110)**

S4 Sample annealed at 600 $^0$C was completely transformed into $RuO_2$ which showed polycrystalline nature with crystal orientation in (110), (200) and (220) crystal direction. The inset image displayed in figure 6 (d) is a magnified view of the $RuO_2$ orientation along the (110) direction as it is closed to substrate orientation of rutile $TiO_2$ (110).[30] There was not any evidence of Ru metal presence in the diffraction pattern as no peak associated with hcp was found. Complete phase transformation was achieved from metallic Ru to $RuO_2$ with the annealing at 600 $^0$C in open air which could be attributed to incorporation of oxygen in the lattice of Ru to form stable polycrystalline $RuO_2$ films. With the annealing in oxygen environment, as deposited Ru metal (sample S1 asdep) changed crystal phase from HCP to rutile $RuO_2$ at annealing temperature of 600 $^0$C ( sample S4).

Ex situ lab based XPS analysis was utilized to understand the chemical state changes of Ru metal films subjected to annealing at different temperatures. The Ru films annealed at 400 $^0$C, 500 $^0$C and 600 $^0$C revealed progressive changes in the chemical composition indicative of Ru oxidation. With the annealing at elevated temperatures, the peaks associated with oxidation state $Ru^{+4}$, emerged at higher binding energy and became more prominent with temperature. **Figure 7** showed the area ratio of lattice oxygen ($O_{lattice}$) to the area ratio of Ru3d calculated for different temperatures. This clearly showed the increase in the ratio as more oxygen was incorporated into the Ru matrix as compared to the as deposited Ru metal. The $Ru^0$ peak was observed at 279.9 eV and another peak for $Ru^{+4}$ was seen at 280.6 eV. The $Ru3d_{3/2}$ doublet was fitted with spin orbit splitting of 4.1 eV as shown in **figure 8 (a).** A dominant peak at 280.6 eV in sample S2 annealed at 400 $^0$C indicates the formation of $Ru^{+4}$ due to oxidation of top layers of as deposited Ru metal film, while a secondary peak at 279.9 eV suggests the presence of $Ru^{(0)}$. In the as-deposited Ru metal (sample S1), the intensity from $Ru^{+4}$ peak was small. This confirms that annealing slowly changed the phase from Ru metal to $RuO_2$ from surface to bulk. Therefore, $O_{lattice}$ proportion was more in sample S2 as compared to as deposited sample S1. As this was done on lab based XPS which uses lower photon energy so probing depth is shallower (e.g., ~6 nm for $RuO_2$ with the Al Kα source used here).[52] However, XRD of sample S2 annealed at 400 $^0$C did not show any orientation for $RuO_2$ as it is showing more bulk properties. Sample S3 annealed at 500 $^0$C showed complete transformation of the film from Ru to $RuO_2$ as binding energy for $Ru3d_{5/2}$ was located at 280.6 eV. The absence of $Ru^{(0)}$ signals confirms further progression of oxide formation further into the bulk of the film. [36]

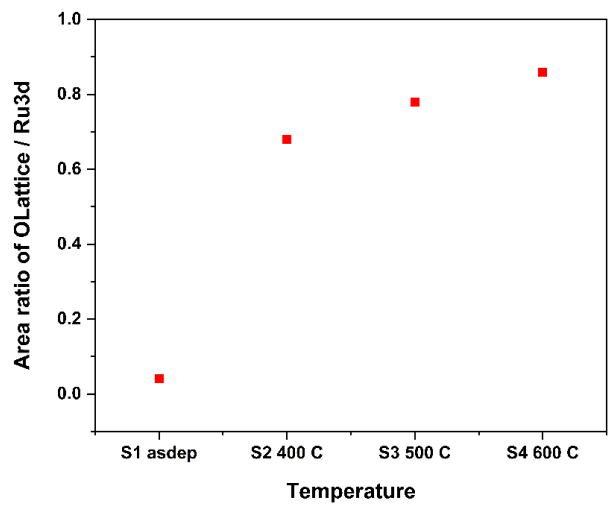

**Figure 7 Area ratio of Olattice to Ru3d of as deposited and annealed samples.**

Concomitantly, satellite peaks became more pronounced and shifted to higher binding energies with the change in oxidation state from $Ru^{(0)}$ to $Ru^{+4}$.[49] These observations indicate a significant transformation in the chemical composition of the films from metal Ru to a more stable oxidized phase of $RuO_2$, particularly evident at the higher annealing temperatures in air. The literature references two additional phases of ruthenium oxide, specifically RuO$_3$ and RuO$_4$. However, experimental evidence for assigning Ru3d$_{5/2}$ binding energies to RuO$_3$ and RuO$_4$ is still in debate. [35, 37] Higher temperatures facilitate the diffusion of oxygen into the ruthenium matrix, promoting the formation of stable ruthenium oxide. The area of both satellite peaks was observed to be increased with an increase in the temperature due to more oxidation. The stoichiometric ratio of O/Ru was increased with temperature which confirms the oxidation of Ru as deposited metal films. **Figure 8 (b)** shows the core level oxygen peak of ruthenium films annealed at different temperatures.

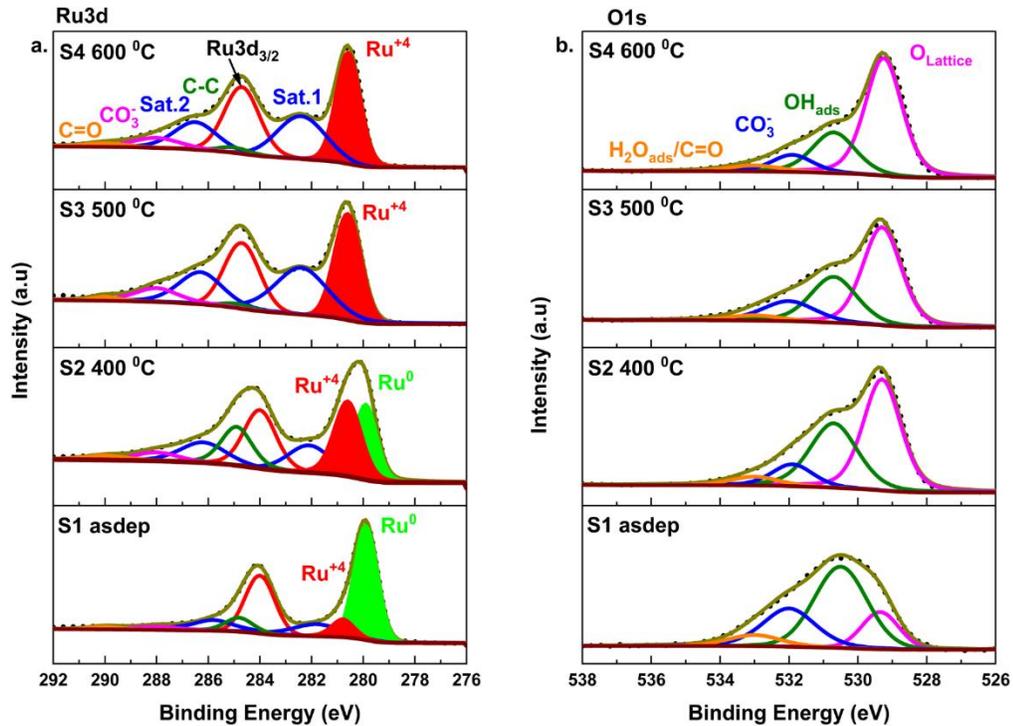

**Figure 8. Core level of (a) Ru3d and (b) O1s spectra of as-deposited Ru (sample S1) to annealed sample S4 at 600 $^0$C from lab based XPS .**

The O1s XPS spectra displayed distinct peaks at the binding energies observed at 529.3 eV, 530.7 eV and 532 eV and 533 eV provide insights into the change of chemical state of oxygen within the annealed samples. The peak at 529.3 eV is typically indicative of oxygen bonded to metal atoms in a metal oxide lattice. The peak at 530.7 eV is assigned to adsorbed hydroxyl species, peak at 532 eV to oxygen-containing carbonate species and The 533 eV peak could be attributed to adsorbed water ($H_2O_{ads}$) and organic carbon (C=O) which overlaps in binding energy. The $O_{Lattice}$ signal emerged as a prominent peak at an annealing temperature of 400 $^0$C, potentially due to the formation of an oxide layer on the surface of the Ru metal.

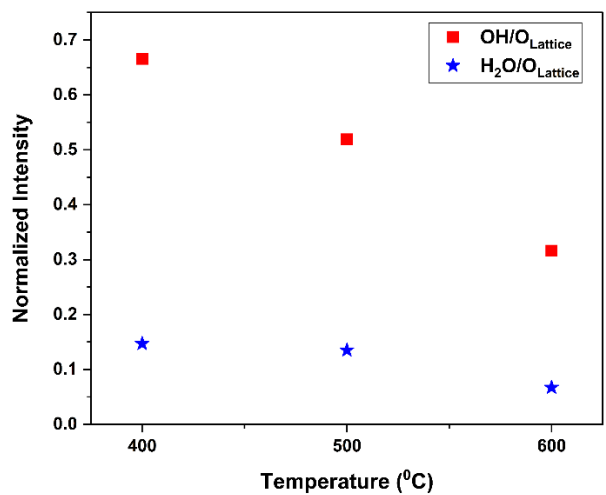

**Figure 9 Annealing temperature dependent trends of -$OH_{ads}$ and $H_2O_{ads}$ concentration with normalized intensity to $O_{Lattice}$.**

**Figure 9** showed the area ratio of $OH_{ads}$ and $H_2O_{ads}$ were normalized to area of $O_{Lattice}$ species which shows reduction in the $OH_{ads}$ and $H_2O_{ads}$ with respect to annealing temperature. The phase transformation observed in the XRD at 600 $^0$C, is confirmed in the XPS studies with the shifting of the spin orbit doublet of Ru3d to the higher binding energy. Thus, lab based XPS shows S2 to be a mixture of $Ru^0$ and $Ru^{+4}$. The Raman spectrum measured on annealed samples in **figure 10** shows promising peaks associated with $RuO_2$ phase. The Raman peak observed at 189 $cm^{-1}$ in the as-deposited ruthenium sample vanished in the sample S4 annealed at 600 $^0C$ as this sample is transformed from metal Ru into $RuO_2$. The peaks observed at 522 $cm^{-1}$, 602 $cm^{-1}$ and 703 $cm^{-1}$ were associated with oxidation peaks of Ruthenium. The peak at 522 $cm^{-1}$ was assigned to the Eg mode associated with the out-of-plane vibration of O atoms. The 602 and 703 $cm^{-1}$ peaks were assigned as the A1g and B2g modes, respectively, that have been associated with vibrations of the two O atoms with respect to the Ru atom. [38,39,40]

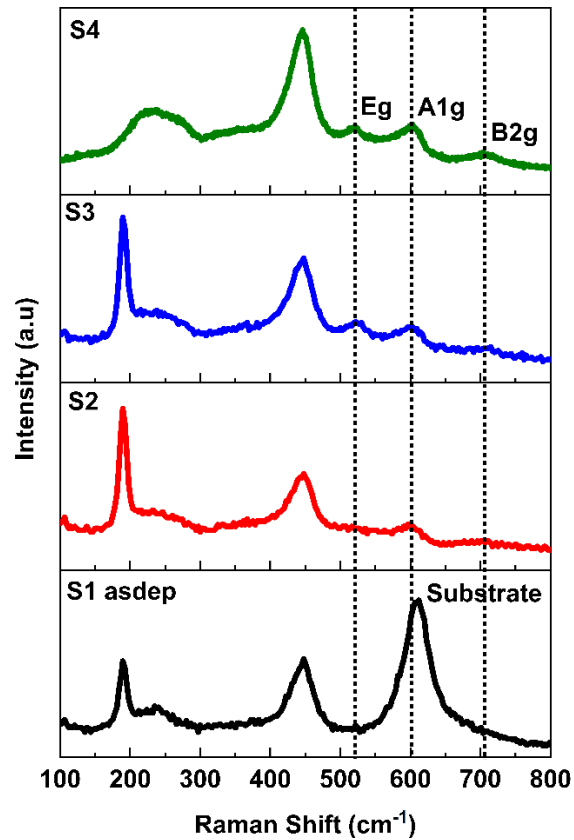

**Figure 10. Raman spectrum of as-deposited sample S1 and annealed samples S2 400 $^0$C, S3 500 $^0$C, S4 600 $^0$C**

## 3.3 Investigations of the interaction of water on as-deposited and annealed Ru films using tender X-ray APXPS

We utilized tender X ray XPS to study the progression of oxide formation from the surface to the bulk due to its ability to probe surface and near-surface regions under realistic conditions, providing insights into the depth-dependent chemical states and oxidation processes. This was done using synchrotron XPS with tender X-rays of 4000 eV energy. The samples were analyzed at water pressures of 1 Torr, 10 Torr, and 17 Torr to approach 100% relative humidity conditions. The samples were exposed to water vapor during measurements to investigate surface reactivity under elevated pressure. The APXPS technique has been employed to investigate the adsorption behavior of hydroxide (OH) and water ($H_2O$) on transition metals and metal oxides, including $TiO_2$, CoOOH, $SrTiO_3$, and $Cu_2O$. [42,43,44,45] Researchers have been using APXPS to investigate the effects of surface chemistry on catalytic activity for $CO_2$ reduction and to examine the electrical potential at the liquid/electrode interface during electrochemical reactions. Here we also studied the effect of change in phase from Ru metal to $RuO_2$ on surface adsorbed species. These observations highlight the crucial role of surface oxidation in enhancing catalytic activity under near-ambient conditions. Here, initially we performed experiments in vacuum with tender X-rays to see the oxidation states on the surface and in the bulk as tender X-rays can penetrate deep into the film. **Figure 11** (a) and (b) represent the Ru3d and O1s fitting results of the as-deposited and annealed samples. As deposited sample S1, sample annealed at 400

$^0$C S2 and annealed at 500 $^0$C S3 showed the presence of Ru$^0$. Tender X-rays have higher penetration depth as compared to lab based XPS (e.g., ~15 nm for RuO$_2$ with the 4000-eV photon energy used here), so XPS studies revealed that there was progressive oxidation of metal Ru film from the surface to bulk.[52] This is why lab based XPS of as-deposited S1 and sample S2 annealed at 400 $^0$C showed mixed phase of metal Ru and RuO$_2$ due to its lower penetration depth. However tender XPS showed as deposited S1, annealed sample S2 at 400 $^0$C and annealed sample S3 at 500 $^0$C mixed oxidation states due to higher penetration depth. After deconvolution of the O1s, a similar approach for peak assignment was adopted as shown in figure 3 for Olattice at 529.3, OH$_{ads}$ at 530.7, metal carbonate species at 532 and for adsorbed water H$_2$O$_{ads}$/C=O at 533 eV. [48] The tender X ray showed higher area ratio of peak at 533 eV associated with carbon contamination and adsorbed water as compared to lab based XPS since tender XPS was done almost after three months from the deposition.

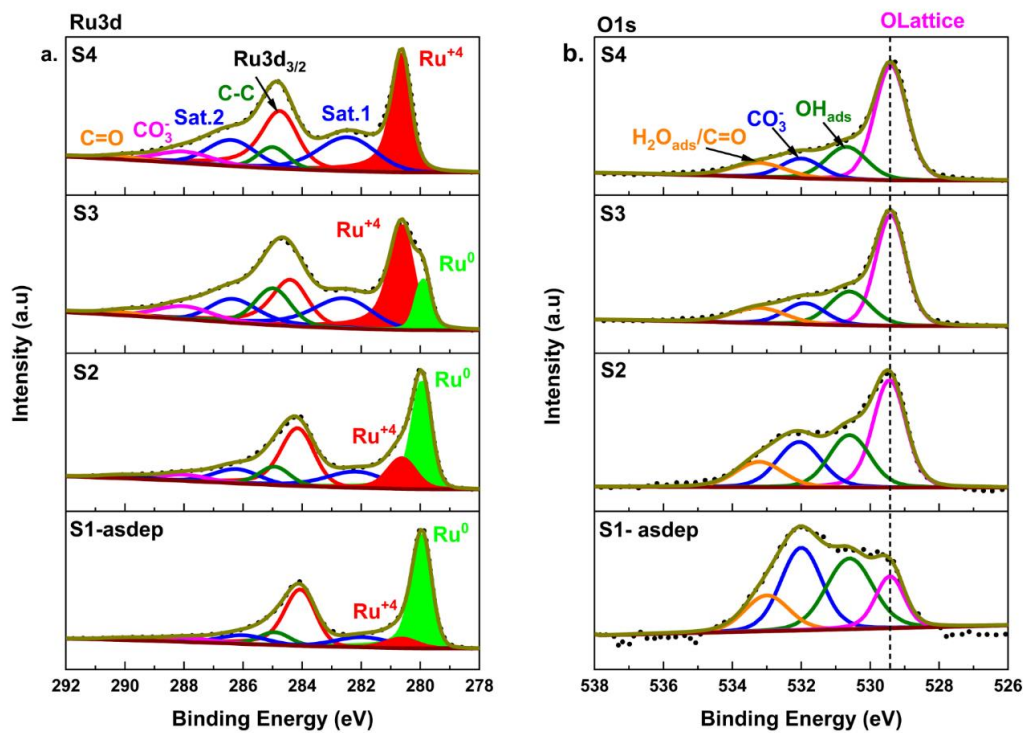

**Figure 11.** Vacuum core level spectra of (a) Ru3d and (b) O1s of S1, S2, S3, and S4 from tender X-ray XPS

**Figure 12 (a)** shows the O1s spectra (normalized to $O_{lattice}$) from the APXPS of as-deposited S1, as the pressure of water in the chamber increased from 1 Torr to 17 Torr, lattice oxygen peak area decreased from 17 to 2% and adsorbed water vapor peak at 533 eV grows from 12 to 66% showing higher affinity towards the water at saturation.

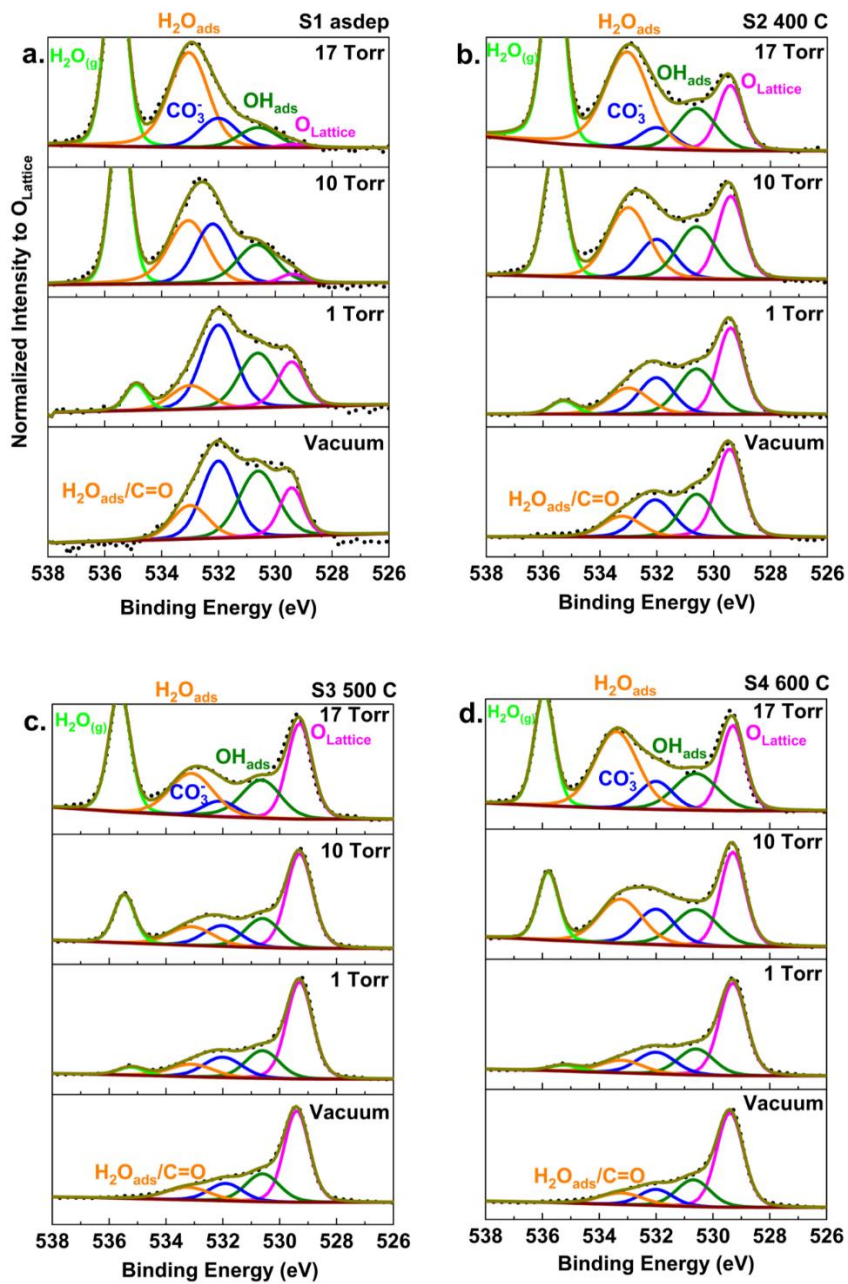

**Figure 12.** O1s APXPS fitting of a) sample S1 asdep, b) sample S2 annealed at 400 $^0$C, c) sample S3 annealed at 500 $^0$C, d) sample S4 annealed at 600 $^0$C from vacuum to 17 Torr

The area from the gaseous water peak ($H_2O_g$) is not taken into consideration while calculating these areas. The concentration of hydroxyl groups decreased from 28 to 13% as pressure was increased, from 1 torr to 17 torr respectively. This reduction in hydroxyl species may be attributed to the lower affinity of Ru metal for hydroxyl groups, as it lacks lattice oxygen that generally promotes dissociative water adsorption.[49] The concentration of adsorbed OH species and adsorbed water species with respect to total area of O1s are decreased for as deposited sample S1, annealed sample S2 at 400 $^0$C and are increased for annealed sample S3 at 500 $^0$C and annealed sample S4 at 600 $^0$C. This observation suggests that lattice oxygen or oxidation state could play a critical role in facilitating surface hydroxylation and water adsorption.[47] **Figure 13 (a) and (b)** show the trend for adsorbed OH normalized to $O_{lattice}$ and adsorbed $H_2O$ normalized to $O_{lattice}$ of samples annealed at 400 $^0$C (S2), annealed at 500 $^0$C (S3) and annealed at 600 $^0$C (S4). Adsorbed OH was found to increase as water may have been adsorbed in dissociative way as annealing temperature is increased. The area ratio of $H_2O_{ads}$ species significantly increased showing the highest affinity of the oxide surface towards water and $H_2O_{ads}$ proportion increased from 11 to 42 % at 17 torr. We observed that as we incorporate more oxygen into Ru metal lattice by annealing in open air, the phase of the Ru metal is transformed into stable $RuO_2$. The enhanced water adsorption with annealing temperature is correlated with increased lattice oxygen which promotes more active surface for adsorption compared to Ru metallic surface. [48]

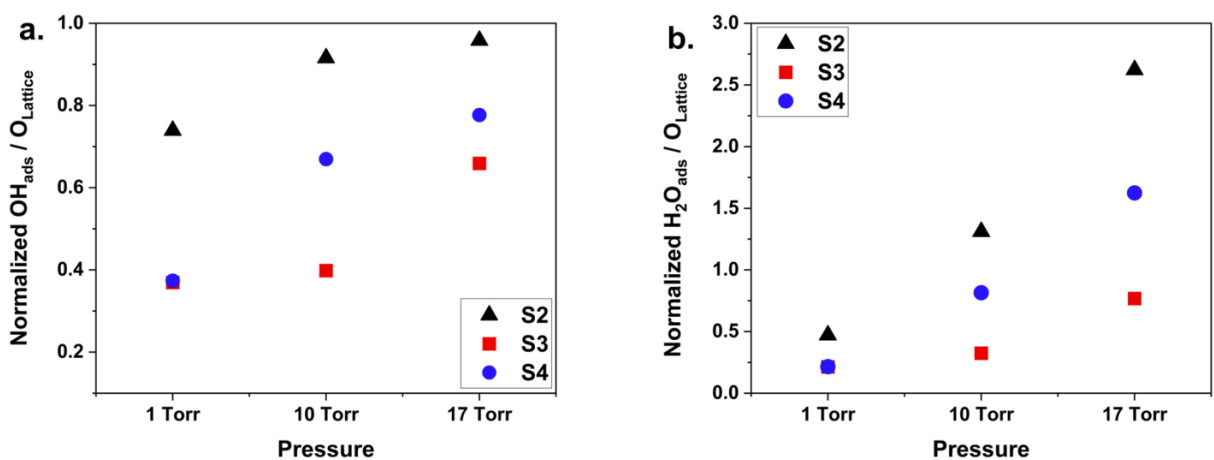

**Figure 13.** Trends of (a) $OH_{ads}$, (b) $H_2O_{ads}$ normalized to lattice oxygen.

**Conclusions**

In this study, uniform as deposited Ru metal film was synthesized at 300 $^0$C by atomic layer deposition and subsequently annealed at elevated temperatures of 400 $^0$C, 500 $^0$C and 600 $^0$C. The structural, morphological, and chemical properties of the as deposited and annealed samples were investigated using lab based and tender X ray XPS, XRD, Raman spectroscopy and AFM. Annealing showed complete phase transformation of Ru metal to $RuO_2$ at 600 $^0$C. The lab based XPS and tender X-ray XPS showed the progressive change in the oxidation states from surface to bulk. Tender X ray XPS vacuum data clearly demonstrated the transformation of Ru metal films into ruthenium oxide, with a prominent emergence of peak at 280.6 eV for $Ru^{+4}$, corresponding to $RuO_2$. There was an increase in grain size and surface roughness when as deposited sample S1 annealed to 400 0C and roughness then reduced with increase in temperature due to relaxation of the film. Raman spectrum showed evidence vibrational modes associated with $RuO_2$. XRD analysis showed that after annealing of the Ru metallic films, it completely transformed into rutile $RuO_2$ at 600 $^0$C. The observed increase in adsorbed species in the APXPS data can be attributed to the change in oxidation state from $Ru^0$ to $Ru^{+4}$, as higher temperatures lead to greater oxygen incorporation into the Ru matrix. Therefore, optimizing the formation of $RuO_2$ surfaces could enhance water adsorption for electrochemical water splitting applications

**Acknowledgement**

The research was supported by the Center for Electrochemical Dynamics and Reactions on Surfaces (CEDARS), an Energy Frontier Research Center, funded by the U.S. Department of Energy (DOE), Office of Science, Basic Energy Sciences (BES) via grant # DE-SC0023415. Part of the work was performed using resources at the Joint School of Nanoscience and Nanotechnology, a member of the National Nanotechnology Coordinated Infrastructure (NNCI), which is supported by the National Science Foundation (Grant ECCS-2025462). The work at the ALS at LBNL was supported by the Director, Office of Science, Office of Basic Energy Sciences, of the US Department of Energy under Contract No. DE-AC02-05CH11231. The work at NILPR was funded by grants of the Romanian Ministry of Scientific Research, Innovation, and Digitalization project PN-III-P4-PCE-2021-1158 and Core Program LAPLAS VII 30N/2023.